\documentclass{article}
\usepackage{spconf,amsmath,graphicx}
\usepackage{multirow, booktabs}
\usepackage{amsmath}
\usepackage{amsfonts}
\usepackage{bm}
\usepackage{enumitem}
 \usepackage{url}
\usepackage[dvipsnames]{xcolor}
 \usepackage{ragged2e}
 \usepackage{multirow}
 \usepackage{pifont}
 \usepackage{multicol}
 \usepackage{caption}
\usepackage{subcaption}
\usepackage{url}
\usepackage{multirow, booktabs}
\usepackage{tabularx}
\newcolumntype{R}{>{\raggedleft\arraybackslash}X}
\newcolumntype{L}{>{\raggedright\arraybackslash}X}
\newcolumntype{C}{>{\centering\arraybackslash}X}
\usepackage{cite}
\usepackage{xpatch,xcolor}
\usepackage{hyperref}
\PassOptionsToPackage{dvipsnames}{xcolor}
\hypersetup{
    colorlinks=true,
    linkcolor=purple,
    filecolor=magenta,      
    urlcolor=purple,
    citecolor=ForestGreen
}
\usepackage{siunitx}
\newcommand{\rev}[1]{#1}

\newcommand{\ablate}[1]{{\color{BrickRed}#1}}

\let\OLDthebibliography\thebibliography
\renewcommand\thebibliography[1]{
  \OLDthebibliography{#1}
  \setlength{\parskip}{.5pt}
  \setlength{\itemsep}{.5pt plus 0.3ex}
}

\newcommand{\cmark}{\ding{51}}
\newcommand{\xmark}{\ding{55}}

\title{Improving Audio Captioning Models with Fine-grained Audio Features, Text Embedding Supervision, and LLM Mix-up Augmentation}

\name{
    \begin{tabular}{c}
    Shih-Lun Wu$^{1}$, Xuankai Chang$^{1}$, Gordon Wichern$^{2}$, Jee-weon Jung$^{1}$,\\ Fran\c{c}ois Germain$^{2}$, Jonathan Le Roux$^{2}$, Shinji Watanabe$^{1}$
    \end{tabular}
    \thanks{This work used \textit{Bridges2-PSC} and \textit{Delta-NCSA} through allocation CIS210014 from the Advanced Cyberinfrastructure Coordination Ecosystem: Services \& Support (ACCESS) program, supported by NSF grants \#2138259, \#2138286, \#2138307, \#2137603, and \#2138296.}}
\address{$^1$ Language Technologies Institute, Carnegie Mellon University, Pittsburgh, PA, USA \\$^2$ Mitsubishi Electric Research Labs (MERL), Cambridge, MA, USA}

\ninept
\begin{document}
\maketitle

\begin{abstract}
Automated audio captioning (AAC) aims to generate informative descriptions for various sounds from nature and/or human activities.
In recent years, AAC has quickly attracted research interest, with state-of-the-art systems now relying on a sequence-to-sequence (seq2seq) backbone powered by strong models such as Transformers. 
Following the macro-trend of applied machine learning research,
in this work, we strive to improve the performance of seq2seq AAC models by extensively leveraging pretrained models and large language models (LLMs).
Specifically, we utilize \textsc{BEATs} to extract fine-grained audio features.
Then, we employ \textsc{Instructor} LLM to fetch text embeddings of captions, and infuse their language-modality knowledge into BEATs audio features via an auxiliary InfoNCE loss function.
Moreover, we propose a novel data augmentation method that uses ChatGPT to produce caption mix-ups (i.e., grammatical and compact combinations of two captions) which, together with the corresponding audio mixtures, increase not only the amount but also the complexity and diversity of training data.
During inference, we propose to employ nucleus sampling and a hybrid reranking algorithm, which has not been explored in AAC research.
Combining our efforts, our model achieves a new state-of-the-art 32.6 SPIDEr-FL score on the Clotho evaluation split, and wins the 2023 DCASE AAC challenge.

\end{abstract}
\section{Introduction}
\label{sec:intro}
Automated audio captioning (AAC) is a multimodal task whose goal is to describe an input audio clip using text.
The descriptions are not restricted to a fixed set of class labels or tags, but are free-form sentences \cite{Drossos_2017_waspaa},
which allow better flexibility and expressivity.
Research progress on AAC has accelerated in recent years thanks to the yearly DCASE challenges, the impressive performance of Transformer-based language models, and the release of the open audio captioning datasets Clotho~\cite{Drossos_2020_icassp} and AudioCaps~\cite{kim2019audiocaps}.
Recent leading works in AAC \cite{yuan2021dcase, xu2022sjtu, yeautomated, chang2023hyu, labbe2023irit, narisetty2021leveraging} all used the sequence-to-sequence (seq2seq) modeling framework,
where an audio encoder is used to extract features from the input audio, and a text decoder learns to generate the caption autoregressively based on the extracted audio features.

While our work is also seq2seq-based,
we extensively leverage machine learning models that are pretrained on large-scale datasets to improve AAC performance from multiple aspects, namely, audio feature extraction, auxiliary training objective, and data augmentation.
We begin with revamping the audio encoder (Section~\ref{subsec:model}).
While PANN \cite{kong2020panns}, a convolution-based audio feature extractor, has long been the tried-and-true choice for AAC research \cite{yuan2021dcase, xu2022sjtu, yeautomated},
many recently proposed audio encoders \cite{chen2022hts, huang2022amae, chen2022beats} have the potential to further improve AAC performance due to more advanced architectures, pretraining objectives, and finer-grained output features.
Specifically, we choose Bidirectional Encoder representation from Audio Transformers (BEATs)~\cite{chen2022beats}, a state-of-the-art multi-label audio tagging model pretrained on AudioSet \cite{gemmeke2017audio}, as our audio encoder.

Next, witnessing the tremendous success of large language models (LLMs) in representing and generating text \cite{ouyang2022training, touvron2023llama},
we use text embeddings from the \textsc{Instructor} Transformer \cite{su2022one} to provide additional supervision (Section~\ref{subsec:instruct}), and employ ChatGPT \cite{chatgpt} to perform a novel mix-up data augmentation (Section~\ref{subsec:chatgpt}).
In previous AAC studies, 
to help linking audio features with concepts in text, which is the output space of AAC tasks, 
\cite{xu2022sjtu} pretrained the PANN encoder with an audio-caption InfoNCE \cite{oord2018representation} contrastive loss, while \cite{yeautomated} used multitask learning to predict keywords in the caption.
In our work, we combine the benefits of both representation learning and multitask training, and also leverage the LLM's rich knowledge of text.
Particularly, we use \textsc{Instructor} to obtain text embeddings for ground-truth captions, and apply an auxiliary InfoNCE loss on a Conformer \cite{gulati2020conformer, miyazaki2020convolution} postencoder to refine/summarize the BEATs audio features and align them with \textsc{Instructor} text embeddings.


For data augmentation, 
other than SpecAugment \cite{park2019specaugment} used commonly in audio-related tasks, researchers have leveraged the original mix-up \cite{zhangmixup} to linearly combine audio/text embeddings from two unrelated samples \cite{labbe2023irit, kouzelis2022_t6a}, synonym substitution on ground-truth captions \cite{chang2023hyu}, and caption concatenation \cite{chang2023hyu}.
As for LLM-based efforts, ChatGPT has been used to compile the large-scale WavCaps \cite{mei2023wavcaps, chang2023hyu, labbe2023irit} AAC dataset by rewriting tags or fragmented descriptions, which are often associated with audio files on the web, to coherent sentences.
Integrating the ideas of mix-up and LLM augmentation methods,
we prompt ChatGPT to mix-up the captions of two audio clips, which produces more natural combined captions than simple text concatenation \cite{chang2023hyu}.
The text mix-ups, when paired with audio mix-ups (i.e., summations of waveforms), increase the amount, complexity, and diversity of our training data.
With all the techniques above, we also discover that nucleus sampling decoding \cite{holtzmancurious} followed by hybrid reranking (Section~\ref{subsec:rerank}) leads to a further performance boost.

Our AAC model attains a state-of-the-art SPIDEr-FL score of 32.6 (on Clotho V2 \cite{Drossos_2020_icassp} evaluation set) and is the winner of the 2023 DCASE AAC Challenge.\footnote{DCASE Challenge 2023 technical report: \scriptsize{\url{https://dcase.community/documents/challenge2023/technical_reports/DCASE2023_Wu_31_t6a.pdf}}.}
Despite the numerous components introduced to our model, we show in our ablation study (Section~\ref{subsec:ablation}) that every component is indispensable to its great performance.
\rev{We open-source our implementation,\footnote{Code: {\scriptsize\url{https://github.com/slSeanWU/beats-conformer-bart-audio-captioner}}.} pretrained model weights,\footnote{Model: {\scriptsize\url{https://huggingface.co/slseanwu/beats-conformer-bart-audio-captioner}}.} and ChatGPT-generated caption mix-ups.\footnote{Mix-ups: {\scriptsize{\url{https://huggingface.co/datasets/slseanwu/clotho-chatgpt-mixup-50K}}.}}}

\begin{figure}[t]
    \centering
    \includegraphics[width=0.85\columnwidth]{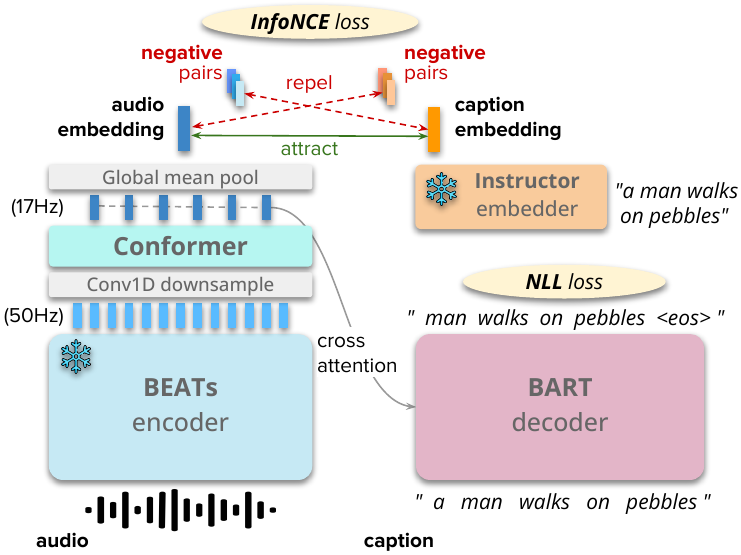}
    \vspace{-3mm}
    \caption{Overview of our Transformer-based captioning system.
    We utilize a frozen BEATs \cite{chen2022beats} to extract audio features from the mel spectrogram.
    On top of BEATs, we attach a Conformer \cite{gulati2020conformer} postencoder to further contextualize the audio features.
    Then, a BART \cite{lewis2020bart} text decoder cross-attends to the contextualized audio features and generates the caption autoregressively.
    To provide text-modality guidance to our encoder stack, we extract the captions' sentence embeddings from an instruction-tuned large language model (LLM), \textsc{Instructor}-XL \cite{su2022one}, and apply an InfoNCE \cite{oord2018representation} auxiliary loss to train Conformer's output audio representation to mimic the corresponding caption's \textsc{Instructor} sentence embedding.}
    \label{fig:overview}
    \vspace{-3mm}
\end{figure}

\section{Method}
\subsection{Network Architecture and Main Loss Function}\label{subsec:model}
We utilize BEATs \cite{chen2022beats} as our main audio encoder.
The BEATs module takes a 16~kHz audio waveform as input, converts the waveform into a mel spectrogram with a 10-millisecond hop size, splits the spectrogram into 2D patches, and transforms the patches into a sequence of representations through 12 self-attention (i.e., Transformer) layers.
Compared to PANN \cite{kong2020panns},
which has been popular in the AAC literature, BEATs comes with some key modifications:
\begin{itemize}[topsep=2pt,itemsep=1pt,leftmargin=*]
    \item \textbf{Architecture:} BEATs features a Transformer backbone, while PANN is based on a convolutional neural network (CNN).
    \item \textbf{Pretraining objectives:} While both BEATs and PANN are pretrained on AudioSet \cite{gemmeke2017audio}, a general-domain, large-scale audio dataset, BEATs is first trained on masked language modeling \cite{devlin2019bert} of tokenized audio, and then on multilabel audio classification. 
    PANN is only trained on the latter.
    \item  \textbf{Resolution:} BEATs provides more fine-grained outputs at 50 Hz, compared to PANN, which has 1-Hz outputs.
\end{itemize}
Due to these differences, and the better performance of BEATs on AudioSet multilabel classification (50.6\% vs.~43.9\% mean average precision), BEATs would likely be a more suitable selection for the AAC task.
In our pilot experiments, we tried either to finetune the BEATs module or to keep it frozen.
Both options led to similar SPIDEr-FL score, so we simply freeze BEATs to reduce computation and memory footprint.

Given that the BEATs module is frozen, to enable further training on the audio features (more details in Section~\ref{subsec:instruct}),
we attach a convolutional downsampling layer, followed by a 2-layer
Conformer \cite{gulati2020conformer} postencoder on top of the BEATs module.
These additional layers further contextualize the audio features, and reduce the text decoder's workload for summarizing the audio features.

Following the recent trend of AAC models \cite{yuan2021dcase, xu2022sjtu}, 
we adopt a 6-layer BART Transformer decoder \cite{lewis2020bart} to generate captions.
We use the default BART text tokenizer with a 50K vocabulary size, and train the BART's weights from scratch.
The BART decoder cross-attends to the Conformer's output representations and self-attends to the historical caption tokens to generate the next caption token autoregressively.
The main loss function to our BEATs-BART captioning model, applied on the BART's output distributions, is the negative log-likelihood (NLL) of audio captions, i.e.,
\begin{equation}
    \mathcal{L}_\mathrm{NLL} = \mathbb{E}_{(\bm{x}, \bm{y}) \in \mathcal{D_{\text{train}}}} \Big[ \sum_{n=1}^{|\bm{y}|} -\log p(y_n|\bm{y}_{1:n-1}; \bm{x}) \Big] \, , \label{eqn:nll}
\end{equation}
where $\mathcal{D_{\text{train}}}$ is the training dataset, $\bm{x}$ is the input audio waveform, $\bm{y}$ is an audio caption, and $y_n$ is the $n^{\text{th}}$ token in the caption.
A schematic overview of our captioning model is depicted in Fig.~\ref{fig:overview}.

\begin{table*}
\scriptsize
\centering 
\caption{Randomly chosen samples of ChatGPT mix-up augmentations.
In general, ChatGPT is able to faithfully reflect all content in the two input captions grammatically, and sometimes exhibits creativity in sentence structuring and vocabulary choice.}\label{table:chatgpt}
\def\arraystretch{1.2}
\vspace{-3mm}
     \begin{tabular}{ p{0.27\textwidth} p{0.27\textwidth} | p{0.385\textwidth} }
     \toprule
    \textbf{Clotho caption \#1} & \textbf{Clotho caption \#2} & \textbf{ChatGPT mix-up} \\
    \hline
        water flowing over some rocks throughout a creek & in the distance fireworks pop and crackle constantly as they are set off & a serene creek babbles over rocks as distant fireworks pop and crackle in celebration \\ \hline
        a muffled object is dragged along a surface in a room that echoes & 
        several dogs barking with many birds making noise in the background &
        dogs bark in a room that echoes while a muffled object is dragged as birds chirp faintly in the background \\ \hline
        a gate squeals as it sways while birds chirp in the background &
        a machine is whirring loudly at first and then slowly shuts off &
        as the gate sways and creaks a nearby machine loudly whirs before slowly powering down amidst chirping birds \\
     \bottomrule
      \end{tabular}
       \vspace{-3mm}
\end{table*}
\subsection{\textsc{Instructor} Embedding Supervision}\label{subsec:instruct}

To infuse text-related knowledge into our audio (i.e., encoder) features,
we leverage an LLM, \textsc{Instructor}-XL
Transformer \cite{su2022one}, to fetch the text embeddings for ground-truth captions and supervise our encoder stack with them.
\textsc{Instructor} is based on a pretrained T5 \cite{raffel2020exploring} text encoder, that is then finetuned using InfoNCE loss \cite{oord2018representation} on a variety of natural language processing (NLP) tasks, such as classification, reranking, summarization, and text quality evaluation, to learn sentence-level text embeddings.
Task- and domain-specific instructions are prepended to the input text as conditions, e.g., ``\textit{Represent the Medicine statement for retrieval:}'', hence the name \textsc{Instructor}.
In the Massive Text Embedding Benchmark (MTEB) \cite{muennighoff2022mteb}, \textsc{Instructor}-XL is the state of the art on text summarization and reranking tasks,\footnote{As of May 2023, when our project was being carried out.} 
which are closely related to audio captioning.

In our use case, we place ``\textit{Represent the audio caption:}'' as the instruction to the (frozen) \textsc{Instructor} to fetch sentence embeddings from ground-truth captions.
On our BEATs-Conformer encoder stack, we perform mean-pooling along the timestep dimension to obtain a single audio embedding for the input waveform.
We denote the audio embedding and the \textsc{Instructor} caption embedding by $\bm{a}$ and $\bm{c}$ respectively.
An auxiliary InfoNCE loss is computed using in-batch negative samples:
\begin{align}
    \mathrm{sim}(\bm{a}, \bm{c}) &= \exp \Big(\frac{\bm{a}^{\top}\bm{c}}{||\bm{a}|| \, ||\bm{c}||} \cdot \frac{1}{\tau}\Big), \label{eqn:cossim} \\ 
    \mathcal{L}_\mathrm{InfoNCE{\textunderscore}a} &= \mathbb{E}_{\mathcal{B} \subset \mathcal{D_\text{train}}} \Big[\sum_{i=1}^{|\mathcal{B}|} - \log \frac{\mathrm{sim}(\bm{a}_i, \bm{c}_i)}{\sum_{j=1}^{|\mathcal{B}|}\mathrm{sim}(\bm{a}_j, \bm{c}_i)} \Big], \\
    \mathcal{L}_\mathrm{InfoNCE{\textunderscore}c} &= \mathbb{E}_{\mathcal{B} \subset \mathcal{D_\text{train}}} \Big[ \sum_{i=1}^{|\mathcal{B}|} - \log \frac{\mathrm{sim}(\bm{a}_i, \bm{c}_i)}{\sum_{j=1}^{|\mathcal{B}|}\mathrm{sim}(\bm{a}_i, \bm{c}_j)} \Big], \\
    \mathcal{L}_\mathrm{InfoNCE} &= 0.5 \cdot (\mathcal{L}_\mathrm{InfoNCE{\textunderscore}a}+ \mathcal{L}_\mathrm{InfoNCE{\textunderscore}c}) \, ,
\end{align}
where $\mathrm{sim(\cdot, \cdot)}$ is the exponentiated temperature-scaled cosine similarity, $\tau$ is the temperature hyperparameter,\footnote{Generally speaking, a higher temperature makes the contrastive objective more challenging, as the distribution is made less peaky. 
We perform a search in $\tau$ = \{0.03, 0.07, 0.2, 0.5, 1.0\} and find $\tau$ = 0.5 works the best.} $\mathcal{B}$ denotes a sampled mini-batch, and $i, j$ index samples in the mini-batch.
The multitask loss $\mathcal{L}$ used to train our model can hence be written as:
\begin{equation}
    \mathcal{L} = \mathcal{L}_\mathrm{NLL} + \alpha\mathcal{L}_\mathrm{InfoNCE} \, ,
\end{equation}
where $\alpha$ is a hyperparameter and we find $\alpha=1$ works well.

\subsection{ChatGPT Mix-up Augmentation}\label{subsec:chatgpt}
As a novel data augmentation measure,
we employ another LLM, ChatGPT \cite{chatgpt}, to `mix-up' \cite{zhangmixup, gong2021psla} pairs of captions in the Clotho dataset,
and create more complex and diverse in-domain training data.
Specifically, we mix-up captions with different corresponding audio clips, rather than two ground-truth captions for the same audio.
The corresponding audio waveforms are also mixed up to ensure consistency between audio and mixed-up captions.

We collect such mix-up augmentations using the public ChatGPT API.
In the prompt, we ask ChatGPT to ``\textit{Generate a mix of the following two audio captions, and keep the generation under 25 words:}'', and then provide it with two randomly sampled captions from Clotho \cite{Drossos_2020_icassp}.
We explicitly limit the number of words to force ChatGPT to be more concise.
We use the FENSE disfluency detector \cite{zhou2022can}
to filter out poor examples.\footnote{Less than 1\% of ChatGPT mix-ups are detected as disfluent.}
Mix-up of audio waveforms is more straightforward:
we follow the algorithm used in WavLM \cite{chen2022wavlm}, scaling the two waveforms to ensure their relative root-mean-square energy is within $\pm$5 dB before adding them together.

Table~\ref{table:chatgpt} displays a few examples of ChatGPT-generated mix-ups.
We try including either 50K or 100K ChatGPT mix-ups, and using 50K yields a better performance.
The API cost for generating 50K mix-ups is roughly \$8.50.

\subsection{Sampling and Reranking}\label{subsec:rerank}
In past AAC research works,
the most commonly used decoding algorithm 
has been beam search \cite{xu2022sjtu, yeautomated, yuan2021dcase}.
However, we find that, after introducing all the techniques in Section~\ref{subsec:model}$\sim$\ref{subsec:chatgpt}, around 1/3 of generations using \textit{nucleus sampling} \cite{holtzmancurious},
which is known to produce more diverse and informative generations than beam search, score higher in terms of SPIDEr-FL than those using beam search.
This reveals the potential advantage of a sampling-then-reranking approach.

To `pick the right sample' with nucleus sampling,
we propose a hybrid reranking algorithm that utilizes again the knowledge of both our learned audio encoder stack and our text decoder.
The two reranking metrics we consider are:
\begin{itemize}[topsep=3pt,itemsep=1pt,leftmargin=*]
    \item \textbf{Caption log-likelihood:} We feed the input waveform $\bm{x}$ and the generated caption $\bm{\hat{y}}$ into our captioning model to directly compute $\log p(\bm{\hat{y}} \,|\, \bm{x}) = \sum_{n=1}^{|\bm{\hat{y}}|} \log p(\hat{y}_n \,|\, \bm{\hat{y}}_{1:n-1}; \bm{x})$ (cf.~Eq.~(\ref{eqn:nll})).
    As the log-likelihood is computed on decoder outputs, we call this \textbf{decoder reranking}.
    \item \textbf{Audio-caption representation similarity:}
    We feed the generated caption $\bm{\hat{y}}$ into the \textsc{Instructor} model to get its text embedding $\bm{\hat{c}}$, and fetch the audio embedding $\bm{a}$ of the input waveform $\bm{x}$ from our trained audio encoder stack. Then, we compute the cosine similarity between the text and audio embeddings, i.e., $(\bm{a}^{\top}\bm{\hat{c}}) \, / \, (||\bm{a}||\,||\bm{\hat{c}}||)$ (cf.~Eq.~(\ref{eqn:cossim})).
    As the representation from the audio encoder is used here, we refer to this as \textbf{encoder reranking}.
\end{itemize}
Candidate captions are ranked by the weighted sum of the two metrics above (with weights tuned on some held-out validation data),
and we return the highest-scoring one as the final predicted caption.
\rev{Our inference process is about three times slower than simple beam search with a beam size of 4.}

\begin{table*}[ht]
\caption{Ablation study on Clotho \cite{Drossos_2020_icassp} evaluation split.
The results demonstrate that all components in our AAC model, i.e., BEATs \cite{chen2022beats} audio encoder (Section~\ref{subsec:model}), 
\textsc{Instructor} \cite{su2022one} sentence embedding supervision (Section~\ref{subsec:instruct}),
ChatGPT \cite{chatgpt} mix-up augmentation (Section~\ref{subsec:chatgpt}),
and nucleus sampling \cite{holtzmancurious} $\,+\,$reranking (Section~\ref{subsec:rerank}),
are beneficial to the performance.
\rev{We highlight the ablated components in \ablate{brown}.}
}\label{tab:ablation}
 \vspace{-2mm}
\scriptsize
\renewcommand\arraystretch{1.0}
\addtolength{\tabcolsep}{-0.2em}
\centering
\begin{tabularx}{\linewidth}{ L  c c c c c   c c c c c}
\toprule
&  \multicolumn{5}{c}{\textbf{Model components}} & \multicolumn{5}{c}{\textbf{Performance metrics} (in \%)}  \\
\cmidrule(lr){2-6}\cmidrule(lr){7-11}
 &  Audio encoder & \textsc{Instructor} emb. & ChatGPT mix-up & Decoding & Reranking & METEOR & CIDEr & SPICE & SPIDEr & SPIDEr-FL \\
\midrule
\textbf{Full model} & BEATs & \cmark & \cmark & Sampling & Hybrid & \textbf{19.3} & \textbf{50.6} & \textbf{14.6} & \textbf{32.6} & \textbf{32.6} \;\;\;\;\;\;\;\: \\ \midrule
\multirow{3}{*}{\shortstack[l]{\textbf{w/o hybrid rerank}\\and/or \textbf{sampling}}} &  BEATs & \cmark & \cmark & Sampling & \ablate{Decoder only} & 18.6 & 45.1 & 13.7 & 29.4 & 29.4 \ablate{\tiny{\textbf{($-$3.2)}}}\\
& BEATs & \cmark & \cmark & Sampling & \ablate{Encoder only} & 18.0 & 43.7 & 13.4 & 28.5 & 28.5 \ablate{\tiny{\textbf{($-$4.1)}}} \\
& BEATs & \cmark & \cmark & \ablate{Beam search} & \ablate{n.a.} & 18.7 & 47.4 & 13.4 & 30.4 & 30.3 \ablate{\tiny{\textbf{($-$2.3)}}} \\ \midrule
\multirow{2}{*}{\textbf{w/o ChatGPT mixup}}  & BEATs & \cmark & \ablate{\xmark} & Sampling & Hybrid & 19.1 & 47.6 & 13.8 & 30.7 & 30.7 \ablate{\tiny{\textbf{($-$1.9)}}} \\
 & BEATs & \cmark & \ablate{\xmark} & \ablate{Beam search} & \ablate{n.a.} & 18.6 & 47.7 & 13.1 & 30.4 & 30.2 \ablate{\tiny{\textbf{($-$2.4)}}} \\ \midrule
\textbf{w/o \textsc{Instructor}}  & BEATs & \ablate{\xmark} & \cmark & \ablate{Beam search} & \ablate{n.a.} & 18.6 & 45.8 & 13.4 & 29.6 & 29.4 \ablate{\tiny{\textbf{($-$3.2)}}} \\ \midrule
 \multirow{2}{*}{\textbf{w/o BEATs}} & \ablate{PANN} & \cmark & \cmark & Sampling & Hybrid & 17.5 & 39.7 & 12.3 & 26.0 & 26.0 \ablate{\tiny{\textbf{($-$6.6)}}} \\
  & \ablate{PANN} & \cmark & \cmark & \ablate{Beam search} & \ablate{n.a.} & 17.4 & 41.6 & 12.2 & 26.9 & 26.7 \ablate{\tiny{\textbf{($-$5.9)}}} \\
\bottomrule
\end{tabularx}
 \vspace{-2mm}
\end{table*}

\begin{table}
\caption{
Performance comparison (on Clotho \cite{Drossos_2020_icassp} evaluation split) with
 top-ranking methods in recent DCASE challenges.
All models presented here are single models, not ensembles.
`RL' indicates the use of reinforcement learning \cite{rennie2017self} to directly optimize CIDEr score.
For fair comparison, best results in each group (i.e., \textit{w/} or \textit{w/o} RL) are \textbf{bold}-faced.
Notice that RL can lead to a heavy punishment on SPIDEr-FL, the new DCASE official metric, due to fluency flaws.
}\label{tab:compare}
 \vspace{-2mm}
\scriptsize
\renewcommand\arraystretch{1.0}
\addtolength{\tabcolsep}{-0.4em}
\centering
\begin{tabularx}{\linewidth}{ L c  c c c c c}
\toprule
&  & \multicolumn{5}{c}{\textbf{Performance metrics} (in \%)}  \\
\cmidrule(lr){3-7}
 & RL & METEOR & CIDEr & SPICE & SPIDEr & SPIDEr-FL \\
\midrule
\textbf{Ours} & \xmark & \textbf{19.3} & \textbf{50.6} & \textbf{14.6} & \textbf{32.6} & \textbf{32.6} \\
\textbf{Labb{\'e} et al., '23} \cite{labbe2023irit} & \xmark & 19.2 & 48.5 & 13.9 & 31.2 & 31.0 \\ 
\textbf{Cho et al., '23} \cite{chang2023hyu} & \xmark & 18.8 & 48.3 & 13.7 & 31.0 & 30.7 \\ 
\textbf{Ye et al., '22} \cite{yeautomated} & \xmark & 17.8 & 44.5 & 12.7 & 28.6  & n.a.\tiny{ ($\leq$28.6)}  \\
\textbf{Xu et al., '22} \cite{xu2022sjtu} & \xmark & 17.9 & 42.1 & 12.7 & 27.4 & n.a.\tiny{ ($\leq$27.4)}  \\ \midrule
\textbf{Cho et al., '23} \cite{chang2023hyu} & \cmark & \textbf{19.5} & \textbf{52.6} & \textbf{14.3} & \textbf{33.5} & \textbf{22.5} \\
\textbf{Ye et al., '22} \cite{yeautomated} & \cmark & 18.5 & 50.3 & 13.2 & 31.7 &  n.a.\tiny{ ($\leq$31.7)}   \\
\textbf{Xu et al., '22} \cite{xu2022sjtu} & \cmark &   18.6 & 50.9 & 12.0 & 31.5 & n.a.\tiny{ ($\leq$31.5)} \\ 
\bottomrule
\end{tabularx}
 \vspace{-2mm}
\end{table}

\section{Experiments and Results}
\subsection{Training and Inference}
We first pretrain the model on the combined dataset of AudioCaps\footnote{We filter out captions with $<$6 words, leading to 35K remaining samples.} \cite{kim2019audiocaps} and 50K ChatGPT mix-ups of samples from the better-curated but smaller Clotho \cite{Drossos_2020_icassp} dataset for 10 epochs (about 13K gradient steps), and then finetune it on Clotho (development split, $\sim$4K samples) for 40 epochs (or 1.2K steps).
Teacher-forcing is applied on the BART decoder inputs.
We adopt the AdamW optimizer with a 2$\,\times\,$10$^{-4}$
learning rate for the `AudioCaps$\,+\,$ChatGPT mix-up' pretraining stage, and 
2$\,\times\,$10$^{-5}$
for the Clotho finetuning stage.

As the Conformer attention (see Section~\ref{subsec:model}) is the primary memory bottleneck due to the long sequence length of audio features, there is a tradeoff between the batch size that can be used and the downsampling rate for the Conv1D layer between our BEATs and Conformer modules---using less downsampling gives the model finer-grained audio features, but a smaller batch size would lead to less reliable gradients and hamper contrastive learning \cite{chen2020simple}.
Through experiments, we settle on the 3x downsampling rate, which allows a batch size of 32 and achieves the best performance.

We train the model on two NVIDIA A100 (40GB) GPUs, and the two training stages take around 6 and 3 hours respectively.
Next-token prediction accuracy on the Clotho validation split is used as the checkpoint selection criterion.

At inference time,
we experiment with generating \{20, 50, 100\} candidate captions per test case with nucleus sampling,\footnote{Nucleus sampling hyperparameters: temperature 0.5; cumulative distribution truncation point, i.e., top-$p$,  0.95.}
and find that generating 50 strikes the best balance between performance gain and compute efficiency.
Additionally, we leverage the FENSE evaluator \cite{zhou2022can} to filter out generations with fluency issues.
We tune the weights of the reranking metrics (see Section~\ref{subsec:rerank}) on the Clotho validation split and eventually pick \{0.3, 0.7\} respectively for decoder and encoder reranking metrics.\footnote{Weights for ablated models (see Table~\ref{tab:ablation}) are separately tuned to be fair.}

\subsection{Evaluation}
The metrics used to evaluate the quality of generated captions are:
METEOR, CIDEr, SPICE, SPIDEr, and SPIDEr-FL.
METEOR and CIDEr are both based on $n$-gram overlap, with the former penalizing fragmentation between the ground-truth and generated captions, and the latter promoting generating informative words by weighting $n$-grams by their TF-IDF scores.
SPICE focuses on the overlap computed on semantic graphs constructed by objects,
object attributes, and relations.
SPIDEr is the simple mean of CIDEr and SPICE, and it had been the official evaluation metric
in DCASE AAC challenges until 2022.
In 2023, the official metric was changed to SPIDEr-FL, which uses FENSE \cite{zhou2022can}, i.e., a BERT-based binary classifier, to penalize the SPIDEr score of disfluent generations by 90\%.

Our evaluation is done on the public `evaluation' split\footnote{DCASE challenge uses the blind `test' split to rank the submissions.} of the Clotho \cite{Drossos_2020_icassp} dataset, which consists of 1,045 samples.
Evaluation results of our full model can be found in the 1$^{\text{st}}$ row of Table~\ref{tab:ablation}.

\subsection{Comparison with Past and Concurrent Works}
We compare our model (i.e., winner of the DCASE 2023 AAC challenge)
to other top performers in the 2022 and 2023 challenges \cite{ xu2022sjtu, yeautomated, chang2023hyu, labbe2023irit}.
We note that while all the best-scoring systems for each participant were ensemble models, we present the metrics for single models for fairness and practicality reasons.\footnote{Ensembles can contain a wildly different \# of models (e.g., 3$\sim$20), and the performance gain can seldom justify the extra compute required.}
The comparison in Table~\ref{tab:compare} shows that our model is state-of-the-art in terms of the new official metric, SPIDEr-FL, and performs competitively on other metrics.
Moreover, while optimizing CIDEr with reinforcement learning has been popular among challenge submissions, the resulting disfluency issues \cite{mei2021encoder} get severely punished on SPIDEr-FL (see 6$^{\text{th}}$ row in Table~\ref{tab:compare}).

\subsection{Ablation Study}\label{subsec:ablation}
To show that every component in our AAC model is indispensable to achieve the best performance,
we conduct a comprehensive ablation study that tries to remove components \textit{one at a time}.
Table~\ref{tab:ablation} presents the results that corroborate the necessity of all of our model components---Each one of them gives at least a 2-point improvement on SPIDEr-FL,\footnote{We cannot perform hybrid reranking with the `\textbf{w/o \textsc{Instructor}}' setting as the audio encoder is not trained to match the caption text embedding.
Thus, we simply use beam search as it outperforms decoder-only reranking.}
with BEATs audio encoder (replacing the popular PANN) being the most crucial one causing a 6-point difference.

Some intriguing additional findings are:
(i) hybrid reranking is required to outperform beam search (see rows 1$\sim$4 in Table \ref{tab:ablation}), suggesting that decoder and encoder reranking methods are strongly complementary and hence should be used together when possible;
(ii) the `sampling$\,+\,$reranking' decoding approach improves performance the most when both BEATs encoder and ChatGPT mix-ups are used (see rows `1 vs.~4', `5 vs.~6', and `8 vs.~9' in Table~\ref{tab:ablation}).
\section{Conclusion and Future Work}
In this work,
we improved audio captioning models from multiple aspects with an extensive use of pretrained models.
We employed the BEATs Transformer to extract more fine-grained audio features.
We then utilized the \textsc{Instructor} text embeddings for multitask learning to provide rich language-modality guidance.
ChatGPT was also leveraged to generate faithful and fluent caption mix-ups which, when paired with the corresponding audio mix-ups,
increased the size, diversity, and complexity of our training data.
Finally, nucleus sampling and hybrid reranking were used to exploit our model's capabilities to the fullest extent.
We accomplished a state-of-the-art 32.6 SPIDEr-FL score and demonstrated via a thorough ablation study that all components are crucial to our model's success.

Future endeavors may explore audio feature extractors that are pretrained with larger amounts of data \cite{elizalde2023clap} or multimodal supervision \cite{imagebind}.
More advanced reinforcement learning methods \cite{dos2021cider} can also be applied to optimize captioning metrics that correlate well with human judgment \cite{hessel2021clipscore} without introducing disfluency issues.

\clearpage
\bibliographystyle{IEEEbib}
\bibliography{refs}
\end{document}